\documentclass[sigconf]{acmart}

\copyrightyear{2026}
\acmYear{2026}
\setcopyright{cc}
\setcctype{by}
\acmConference[MSR '26]{23rd International Conference on Mining Software Repositories}{April 13--14, 2026}{Rio de Janeiro, Brazil}
\acmBooktitle{23rd International Conference on Mining Software Repositories (MSR '26), April 13--14, 2026, Rio de Janeiro, Brazil}
\acmPrice{}
\acmDOI{10.1145/3793302.3793583}
\acmISBN{979-8-4007-2474-9/2026/04}

\usepackage{booktabs}
\usepackage{multirow}
\usepackage{xcolor}
\usepackage{adjustbox}
\usepackage{colortbl}
\usepackage{balance}
\usepackage{tcolorbox}
\usepackage{hyperref}

\definecolor{ored}{HTML}{FB8957}
\definecolor{dkblue}{HTML}{8AAFC0}

\newtcolorbox{quotebox}{colback=ored!50,boxrule=0pt,colframe=black,fonttitle=\bfseries,before skip=2pt, after skip=2pt, left=2pt, right=2pt, top=2pt, bottom=2pt}

\newtcolorbox{takeawaybox}{colback=dkblue!50,boxrule=0pt,colframe=black,fonttitle=\bfseries,before skip=2pt, after skip=2pt, left=2pt, right=2pt, top=2pt, bottom=2pt}

\begin{document}

\title{Analyzing Message-Code Inconsistency in AI Coding Agent-Authored Pull Requests}

\author{Jingzhi Gong}
\orcid{0000-0003-4551-0701}
\affiliation{%
  \institution{King's College London}
  \city{London}
  \country{UK}
}
\email{jingzhi.gong@kcl.ac.uk}

\author{Giovanni Pinna}
\orcid{0000-0001-8268-3447}
\affiliation{%
  \institution{University of Trieste}
  \city{Trieste}
  \country{Italy}
}
\email{giovanni.pinna@units.it}

\author{Yixin Bian}
\orcid{0000-0001-8569-7107}
\affiliation{%
  \institution{Harbin Normal University}
  \city{Harbin}
  \country{China}
}
\email{yixin.bian@hrbnu.edu.cn}

\author{Jie M. Zhang}
\orcid{0000-0003-0481-7264}
\affiliation{%
  \institution{King's College London}
  \city{London}
  \country{UK}
}
\authornote{Corresponding author.}
\email{jie.zhang@kcl.ac.uk}

\begin{abstract}
Pull request (PR) descriptions generated by AI coding agents are the primary channel for communicating code changes to human reviewers. However, the alignment between these messages and the actual changes remains unexplored, raising concerns about the trustworthiness of AI agents. To fill this gap, we analyzed 23,247 agentic PRs across five agents using PR message-code inconsistency (PR-MCI). We contributed 974 manually annotated PRs, found 406 PRs (1.7\%) exhibited high PR-MCI, and identified eight PR-MCI types, revealing that \emph{descriptions claim unimplemented changes} was the most common issue (45.4\%). Statistical tests confirmed that high-MCI PRs had 51.7\% lower acceptance rates (28.3\% vs. 80.0\%) and took 3.5$\times$ longer to merge (55.8 vs. 16.0 hours). Our findings suggest that unreliable PR descriptions undermine trust in AI agents, highlighting the need for PR-MCI verification mechanisms and improved PR generation to enable trustworthy human-AI collaboration.
\end{abstract}

\ccsdesc[500]{Software and its engineering~Collaboration in software development}

\keywords{LLMs, Code Review, Human-AI Collaboration, AI Trustworthiness, Empirical SE, Mining Software Repositories, LLM4Code, AI4SE}

\maketitle

\vspace{-0.1cm}
\section{Introduction}
\label{sec:introduction}

AI coding agents are increasingly acting as autonomous teammates in software development~\cite{watanabe2025use, wang2025ai, gong2025ga4gc}, generating code and opening pull requests (PRs) with natural-language descriptions that communicate intent~\cite{ogenrwot2025patchtrack}. Unlike human-written PRs, AI agent-authored PRs (Agentic-PRs) are largely black boxes~\cite{watanabe2025use}, forcing reviewers to rely on PR descriptions to understand changes and assess correctness~\cite{zhang2022pull, ford2019beyond}. As a result, the reliability of PR descriptions is critical for enabling effective human-AI collaboration.

However, AI-generated descriptions (title + body) are not always faithful to the underlying code, as generative models can produce hallucinated or incorrect statements~\cite{huang2025survey, twist2025library}, undermining trust in AI agents. While message-code inconsistency (MCI) has been studied for human-written commit messages~\cite{dong2023revisiting, zhang2025codefuse, wen2019large}, little is known about how often Agentic-PR descriptions misalign with code changes, what inconsistency types occur, or whether such misalignment affects reviewer trust and PR outcomes.

To address this gap, we conducted an empirical study analyzing 23,247 Agentic-PRs from the AIDev dataset~\cite{aidev2025} using \emph{PR message-code inconsistency} (PR-MCI) and made the following contributions:

(1) \textbf{Annotated dataset}: We release 974 manually annotated PRs (432 partial/misaligned cases), supporting the development of PR-MCI detection methods and more reliable AI coding agents.

(2) \textbf{RQ1: How often do Agentic-PR descriptions misalign with their code changes?} We found 406 high-MCI PRs (1.7\%) with a 20-fold variation across agents, suggesting that a non-trivial fraction of PRs could mislead reviewers.
    
(3) \textbf{RQ2: What types of message-code inconsistencies occur in AI-generated PRs?} We identified eight PR-MCI types, finding that \emph{Phantom Changes} (descriptions claim unimplemented changes) dominated (45.4\%).
    
(4) \textbf{RQ3: Is message-code inconsistency associated with PR acceptance and review effort?} We showed that high-MCI PRs had 51.7\% lower acceptance rates (28.3\% vs. 80.0\%) and took 3.5$\times$ longer to merge (55.8 vs. 16.0 hours).

Overall, our findings indicate that while AI agents can generate PRs at scale, the reliability of their descriptions varies substantially and is associated with reviewer trust and PR outcomes, offering actionable insights for developers, tool builders, and researchers.

\vspace{-0.1cm}

\vspace{-0.1cm}
\section{Background and Related Work}
\label{sec:background}

\textbf{Agentic-PRs.} 
AI coding agents are increasingly capable of autonomously authoring and submitting PRs in real-world software projects~\cite{ogenrwot2025patchtrack, brookes2025evolving, watanabe2025use}. In these \emph{Agentic-PRs}, natural-language descriptions play a critical role in communicating intent and scope to human reviewers. However, generative AI systems are known to produce hallucinated or incorrect statements~\cite{huang2025survey, twist2025library}, raising concerns about the reliability and trustworthiness of Agentic-PRs.

\textbf{Message-Code Inconsistency.} 
Commit messages and PR descriptions serve as documentation of developer intent. When these descriptions misalign with actual code changes, they create \emph{message-code inconsistency}~\cite{wang2025hard, dong2023revisiting}. Recent work has studied MCI in human-written commit messages and benchmarked LLMs for detecting text-code inconsistencies~\cite{dong2023revisiting, zhang2025codefuse}, highlighting challenges in generating faithful change descriptions.

\textbf{Gap in PR-MCI.} 
While MCI has been studied for human-written commit messages~\cite{dong2023revisiting,zhang2025codefuse}, and prior work has analyzed PR text for quality concerns~\cite{karmakar2022experience}, very few prior works examine inconsistency in Agentic-PR descriptions\footnote{Note that PR descriptions differ from commit messages: PR descriptions may span multiple commits and serve as the primary communication channel for reviewers.}. This gap is critical because agents generate descriptions automatically without human oversight, potentially producing misleading explanations at scale. PR decision-making research shows that description quality is associated with acceptance rates~\cite{zhang2022pull}, and automatically created PRs can differ in how maintainers interact with them~\cite{wyrich2021bots}. Our work extends this research by examining how description-code consistency is associated with outcomes for Agentic-PRs.

\section{Methodology and Experiment Setup}
\label{sec:methods}

\begin{figure}[!t]
    \centering
    \includegraphics[width=\linewidth]{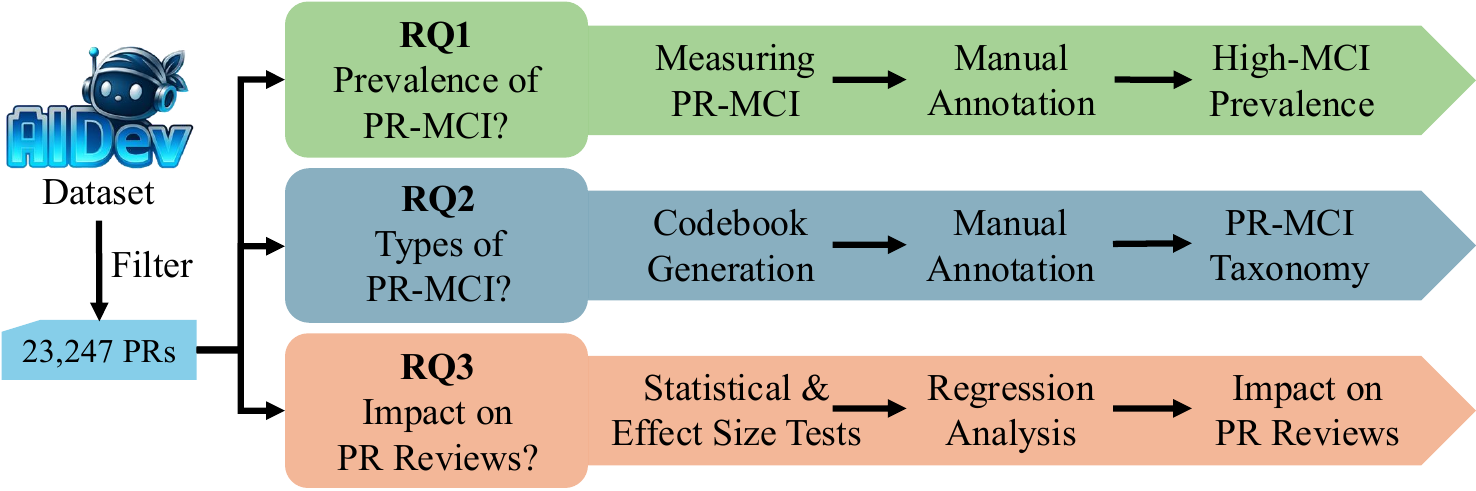}
    \vspace{-0.6cm}
    \caption{Methodology workflow for PR-MCI analysis.}
    \vspace{-0.3cm}
    \label{fig:workflow}
\end{figure}


Figure~\ref{fig:workflow} illustrates our workflow.
We used the \textit{AIDev-pop} subset (33,596 PRs from 2,807 repositories with $>$100 stars) from the \textit{AIDev} dataset \cite{aidev2025}. After filtering for closed PRs, permissive licenses (MIT or Apache~2.0), and six most common task types (95.2\%), the final dataset contains \textbf{23,247 PRs} authored by five AI agents. 

\subsection{Measuring PR-MCI (RQ1)}

Following prior MCI research~\cite{wen2019large, zhang2025codefuse}, we used PR-MCI as the degree to which a PR description (title + body) diverges from the underlying code changes.
We measured PR-MCI using a \textbf{heuristic similarity score} that combines three complementary signals: \emph{scope adequacy} $s_s$ (whether description verbosity matches code churn), \emph{file-type consistency} $s_f$ (whether mentioned file types, e.g., tests or documentation, are actually modified), and \emph{task-type alignment} $s_t$ (whether description language matches the labeled task type).
The similarity score $s \in [0,1]$ aggregates these signals using a weighted scheme~\cite{kaliszewski2016simple}: 
$s = 0.3 \cdot s_s + 0.4 \cdot s_f + 0.3 \cdot s_t$,
where $s_f$ receives the highest weight (40\%) for providing the most concrete and verifiable evidence of misalignment (e.g., claiming test changes when no test files are modified), while $s_s$ and $s_t$ each receive 30\% as complementary but less definitive signals~\cite{barnett2016relationship}
\footnote{Component ablation analysis is available in our \href{https://github.com/gjz78910/PR-MCI/blob/main/supplementary_material/ablation_study.md}{\textcolor{blue}{supplementary material}}.}. 

To validate the PR-MCI score and calibrate the decision threshold, two authors \textbf{manually annotated} 600 PRs, stratified by agent, task type, and score bins, as \emph{aligned}, \emph{partially aligned}, or \emph{misaligned}, achieving strong agreement ($\kappa=0.892$). PRs with similarity below $\theta=0.61$ are labeled \emph{high-MCI}, where $\theta$ is selected by optimizing F1 on labeled validation data~\cite{fan2007study} after conducting sensitivity analysis across threshold ranges \cite{lipton2014optimal}; five-fold cross-validation shows stable threshold selection (mean=0.606, std=0.008) \cite{zhang2025codefuse}. We also evaluated alternative methods: an embedding model (Qwen3-Embedding-0.6B~\cite{qwen3_embedding_06b_2024}) performs poorly (F1=0.150), and an agreement method (both methods must agree) achieves lower F1 (0.567 vs. 0.630), indicating both are insufficient (see Table~\ref{tab:prevalence}).
Consistent with large-scale inconsistency studies~\cite{wen2019large}, we report prevalence with \textbf{95\% Wilson confidence intervals}.

\subsection{PR-MCI Taxonomy Development (RQ2)}

Following mixed-method inconsistency studies that derive taxonomies from manually coded, automatically retrieved candidates \cite{wen2019large}, we developed a PR-MCI taxonomy through an \textbf{iterative process}: GPT-5.2~\cite{openai_gpt52_2025} first generates an initial codebook with eight categories, which two annotators refine through discussion before \textbf{manually classifying} 432 partial/misaligned PRs from 974 PRs (600 validation + 374 additional high-MCI PRs from RQ1).

Note that RQ1 and RQ3 use a \emph{strict} binary definition of high-MCI (treating \emph{partially aligned} PRs as aligned) to identify only clearly misaligned PRs, whereas RQ2 includes both partial and misaligned PRs to capture a broader range of inconsistency patterns for a more comprehensive taxonomy. As in prior work, the taxonomy reflects the analyzed subset and may not capture all inconsistency types.

\subsection{Statistical Tests (RQ3)}

To examine associations between PR-MCI and outcomes, we compared high- and low-MCI PRs using {standard statistical tests}. Following PR decision research~\cite{zhang2022pull}, we analyzed acceptance rate, time to merge, review count, and comment count. For binary outcomes (acceptance), we used \textbf{Chi-square tests} \cite{mchugh2013chi} with \textbf{Cramér's~V} \cite{cramer1999mathematical}; for continuous variables (e.g. time), \textbf{Mann-Whitney~U tests} \cite{mcknight2010mann} with \textbf{Cliff's~$\delta$} \cite{macbeth2011cliff}. Significance is assessed at $p<0.05$.

To control for confounders, we fitted \textbf{regression models} including log-transformed code churn, files changed, task type, and agent, following established PR outcome research~\cite{zhang2022pull}.

\section{Results}
\label{sec:results}

\subsection{RQ1: Prevalence of PR-MCI}
\label{sec:results-rq1}

\begin{table}[t]
\centering
\caption{High-MCI Prevalence and Validation Metrics. [lower, upper] show 95\% Wilson confidence intervals. \setlength{\fboxsep}{1.5pt}\colorbox{ored!50}{Highlighted} values are from the primary Heuristic MCI scoring method.}
\vspace{-0.4cm}
\label{tab:prevalence}
\begin{adjustbox}{width=\columnwidth,center}
\begin{tabular}{p{3cm}rrr}
\toprule
\textbf{Category} & \textbf{Heuristic (PRIMARY)} & \textbf{Embedding} & \textbf{Agreement} \\
\midrule
Overall Prevalence & \cellcolor{ored!50}1.7\% [1.6, 1.9] & 60.0\% & 1.4\% \\
\hline
\hline
\multicolumn{4}{c}{\textbf{High-MCI Prevalence by Agent (\%)}} \\ 
        \hline
GitHub Copilot & \cellcolor{ored!50}8.7\% [7.7, 9.9] & 83.5\% & 7.1\% \\
Cursor & \cellcolor{ored!50}4.5\% [3.2, 6.4] & 76.8\% & 3.4\% \\
Claude Code & \cellcolor{ored!50}0.9\% [0.2, 3.3] & 74.5\% & 0.9\% \\
OpenAI Codex & \cellcolor{ored!50}0.8\% [0.6, 0.9] & 52.7\% & 0.6\% \\
Devin & \cellcolor{ored!50}0.4\% [0.2, 0.7] & 75.5\% & 0.2\% \\
\hline
\hline
\multicolumn{4}{c}{\textbf{High-MCI Prevalence by Task Type (\%)}} \\ 
\hline
Chore & \cellcolor{ored!50}4.0\% [2.7, 5.9] & 54.5\% & 3.0\% \\
Refactor & \cellcolor{ored!50}3.5\% [2.7, 4.5] & 47.7\% & 2.2\% \\
Bug Fix & \cellcolor{ored!50}2.1\% [1.8, 2.5] & 49.5\% & 1.6\% \\
Feature & \cellcolor{ored!50}1.5\% [1.3, 1.8] & 69.7\% & 1.3\% \\
Documentation & \cellcolor{ored!50}1.0\% [0.7, 1.5] & 57.4\% & 0.8\% \\
Test & \cellcolor{ored!50}1.0\% [0.6, 1.5] & 49.3\% & 0.6\% \\
\hline
\hline
\multicolumn{4}{c}{\textbf{Validation Metrics (n=600)}} \\
\hline
Precision & \cellcolor{ored!50}0.742 & 0.083 & 0.760 \\
Recall & \cellcolor{ored!50}0.548 & 0.786 & 0.452 \\
F1 & \cellcolor{ored!50}0.630 & 0.150 & 0.567 \\
\bottomrule
\end{tabular}
\end{adjustbox}
\end{table}

\begin{figure}[!t]
    \centering
    \includegraphics[width=\linewidth]{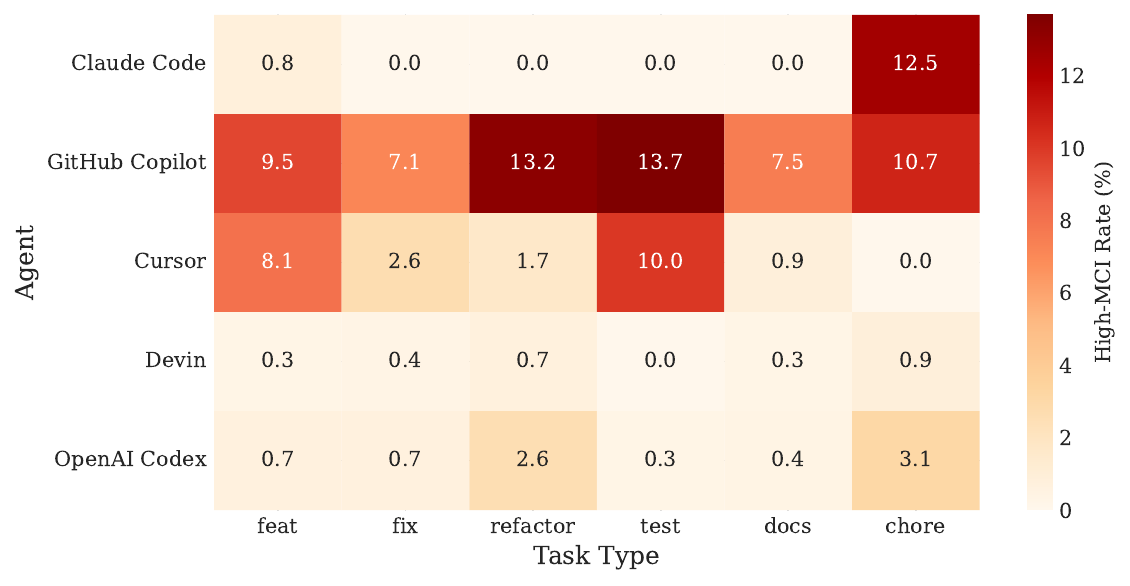}
    \vspace{-0.7cm}
    \caption{Heatmap of high-MCI prevalence (\%) by agent$\times$task.}
    \vspace{-0.3cm}
    \label{fig:rq1_heatmap}
\end{figure}

Table~\ref{tab:prevalence} reports prevalence and validation metrics.
Overall, \textbf{406 out of 23,247 Agentic-PRs (1.7\%)} exhibited high PR-MCI (similarity score below 0.61), where descriptions poorly match code changes, suggesting that a small but non-trivial fraction could mislead reviewers. The prevalence revealed a \textbf{20-fold difference} across agents: GitHub Copilot had the highest rate (\textbf{8.7\%}, 234 high-MCI PRs out of 2,675), followed by Cursor (4.5\%, 31 out of 682).\footnote{Sensitivity analysis (threshold 0.55-0.65) suggested that relative ordering remained consistent across thresholds, as in our \href{https://github.com/gjz78910/PR-MCI/blob/main/supplementary_material/sensitivity_analysis.md}{\textcolor{blue}{supplementary material}}.}

Across task types, we observed a \textbf{4-fold variation}: \textbf{chore PRs} had the highest inconsistency rate (\textbf{4.0\%, 24 out of 600}), followed by refactoring (3.5\%, 54 out of 1,553) and bug fixes (2.1\%, 113 out of 5,319). Features (1.5\%), documentation (1.0\%), and test PRs (1.0\%) showed the lowest rates. The heatmap (Figure~\ref{fig:rq1_heatmap}) reveals agent$\times$task interactions: GitHub Copilot showed elevated rates across most task types, particularly for test (13.7\%) and refactor tasks (13.2\%).

\begin{quotebox}
\noindent
\textbf{RQ1:} \textbf{406 PRs (1.7\%)} exhibited high PR-MCI, suggesting that a non-trivial fraction of PRs could mislead reviewers. In particular, GitHub Copilot showed 8.7\% (20x higher than Devin's 0.4\%), and chore tasks showed 4.0\% (4x higher than test tasks' 1.0\%).
\end{quotebox}


\subsection{RQ2: PR-MCI Taxonomy}
\label{sec:results-rq2}

\begin{figure*}[!t]
    \centering
    \includegraphics[width=\linewidth]{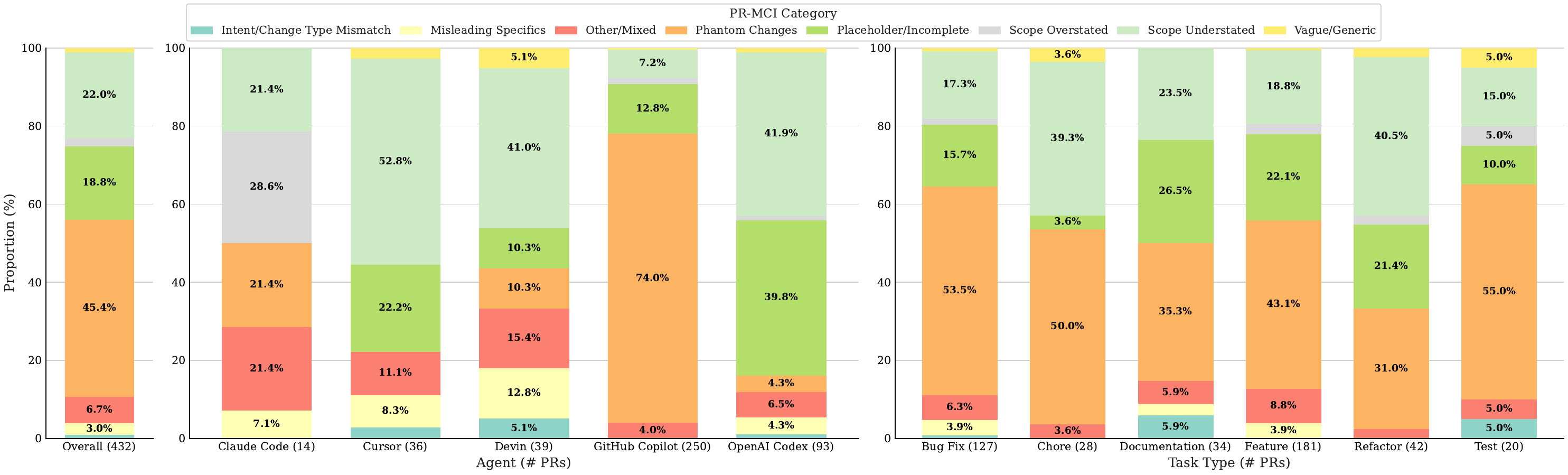}
    \vspace{-0.75cm}
    \caption{Distribution of PR-MCI categories overall, by coding agent, and by task type.}
    \vspace{-0.3cm}
    \label{fig:rq2_category_combined}
\end{figure*}

Figure~\ref{fig:rq2_category_combined} shows the taxonomy over all, by agent, and by task type. In particular, the most common PR-MCI type was \textbf{Phantom Changes} (\textbf{45.4\%}), where PR descriptions claim unimplemented changes. This was followed by Scope Understated (22.0\%), where descriptions omit significant changes, and Placeholder/Incomplete (18.8\%), which used generic or boilerplate text\footnote{Examples of different PR-MCI types are provided in our \href{https://github.com/gjz78910/PR-MCI/blob/main/supplementary_material/taxonomy_definitions.md}{\textcolor{blue}{supplementary material}}.}. 

For agent-specific patterns, GitHub Copilot was dominated by Phantom Changes (74.0\%), while Scope Understated was the primary issue for Cursor (52.8\%) and Devin (41.0\%). OpenAI Codex showed a balanced distribution with Placeholder/Incomplete (39.8\%) and Scope Understated (41.9\%). 

Task-specific patterns showed that Test (55\%) and Bug Fix (53.5\%) PRs were most prone to Phantom Changes, while Refactor PRs showed Scope Understated (40.5\%) as the dominant issue.


\begin{quotebox}
\noindent
\textbf{RQ2:} \textbf{Phantom Changes} was the dominant PR-MCI type (45.4\%), followed by Scope Understated (22.0\%) and Placeholder/Incomplete (18.8\%), suggesting that AI agents frequently claim unimplemented changes or omit significant changes.
\end{quotebox}

\subsection{RQ3: Impact on Review Outcomes}
\label{sec:results-rq3}

\begin{table}[t]
\centering
\caption{Review and Outcome Metrics by MCI Level. \setlength{\fboxsep}{1.5pt}\colorbox{ored!50}{Highlighted} values indicate statistically significant differences between High-MCI and Low-MCI PRs with non-negligible effect sizes. See \S\ref{sec:methods} for statistical test details.}
\vspace{-0.2cm}
\label{tab:summary_stats}
\begin{adjustbox}{width=\columnwidth,center}
\begin{tabular}{p{2.3cm}rrrp{1.25cm}rr}
\toprule
\textbf{Metric} & \textbf{Low-MCI} & \textbf{High-MCI} & \textbf{$p$-value} & \textbf{Effect~Size} & \textbf{\# Low} & \textbf{\# High} \\
\midrule
Acceptance~Rate~(\%) & 80.0 & \cellcolor{ored!50}28.3 & \textbf{$<$0.001} & \textbf{Small} & 22,841 & 406 \\
Mean~Time~to~Merge~(h) & 16.0 & \cellcolor{ored!50}55.8 & \textbf{$<$0.001} & \textbf{Small} & 18,282 & 115 \\
Mean~\#~Review & 0.74 & 0.42 & \textbf{0.039} & Negligible & 22,841 & 406 \\
Mean~\#~Comment & 0.64 & 0.37 & \textbf{0.012} & Negligible & 22,841 & 406 \\
\hline
\hline
\multicolumn{7}{c}{\textbf{Acceptance Rate by Agent (\%)}} \\ 
\hline
Claude Code & 69.3 & 50.0 & 1.000 & Negligible & 218 & 2 \\
GitHub Copilot & 59.4 & \cellcolor{ored!50}3.4 & \textbf{$<$0.001} & \textbf{Medium} & 2,441 & 234 \\
Cursor & 73.3 & 67.7 & 0.638 & Negligible & 651 & 31 \\
Devin & 58.3 & 45.5 & 0.580 & Negligible & 2,880 & 11 \\
OpenAI Codex & 87.2 & 62.5 & \textbf{$<$0.001} & Negligible & 16,651 & 128 \\
\hline
\hline
\multicolumn{7}{c}{\textbf{Acceptance Rate by Task Type (\%)}} \\
\hline
Chore & 80.2 & \cellcolor{ored!50}45.8 & \textbf{$<$0.001} & \textbf{Small} & 576 & 24 \\
Documentation & 89.8 & \cellcolor{ored!50}34.5 & \textbf{$<$0.001} & \textbf{Small} & 2,786 & 29 \\
Feature & 80.1 & \cellcolor{ored!50}27.7 & \textbf{$<$0.001} & \textbf{Small} & 10,782 & 166 \\
Bug Fix & 73.6 & \cellcolor{ored!50}17.7 & \textbf{$<$0.001} & \textbf{Small} & 5,206 & 113 \\
Refactor & 78.1 & \cellcolor{ored!50}42.6 & \textbf{$<$0.001} & \textbf{Small} & 1,499 & 54 \\
Test & 84.3 & \cellcolor{ored!50}25.0 & \textbf{$<$0.001} & \textbf{Small} & 1,992 & 20 \\
\hline
\hline
\multicolumn{7}{c}{\textbf{Time to Merge (h) by Agent (merged PRs only)}} \\
\hline
Claude Code & 31.4 & 6.1 & 0.817 & Negligible & 149 & 1 \\
GitHub Copilot & 70.2 & 24.0 & 0.163 & \textbf{Small} & 1,420 & 8 \\
Cursor & 23.1 & 28.9 & 0.333 & Negligible & 477 & 21 \\
Devin & 21.8 & 14.4 & 0.976 & Negligible & 1,669 & 5 \\
OpenAI Codex & 5.7 & \cellcolor{ored!50}38.6 & \textbf{$<$0.001} & \textbf{Medium} & 14,509 & 77 \\
\hline
\hline
\multicolumn{7}{c}{\textbf{Time to Merge (h) by Task Type (merged PRs only)}} \\
\hline
Chore & 16.5 & 81.4 & 0.181 & \textbf{Small} & 460 & 11 \\
Documentation & 8.9 & 1.1 & 0.140 & \textbf{Small} & 2,493 & 10 \\
Feature & 11.3 & \cellcolor{ored!50}16.4 & \textbf{$<$0.001} & \textbf{Medium} & 8,617 & 44 \\
Bug Fix & 20.6 & \cellcolor{ored!50}34.4 & \textbf{$<$0.001} & \textbf{Medium} & 3,812 & 20 \\
Refactor & 15.5 & \cellcolor{ored!50}51.2 & \textbf{0.012} & \textbf{Small} & 1,165 & 22 \\
Test & 6.4 & 80.6 & 0.825 & Negligible & 1,677 & 5 \\
\bottomrule
\end{tabular}
\end{adjustbox}
\end{table}

Table~\ref{tab:summary_stats} reports summary statistics for key metrics with statistical tests and breakdowns by agent and task type.
Notably, high-MCI PRs were associated with a significantly lower acceptance rate: \textbf{28.3\%} vs. \textbf{80.0\%}, a difference of 51.7\% ($p < 0.001$). The corresponding effect size was small (Cramér's V = 0.166), indicating a non-trivial practical difference. The impacts on acceptance rate varied substantially by agent: for GitHub Copilot, the effect was largest (3.4\% vs. 59.4\%, a 55.9 percentage point drop, $p < 0.001$). The effect of high-MCI was even more obvious by task type, with statistically significant differences for all tasks (all $p < 0.001$ with small effect sizes).

For the time to merge, high-MCI PRs took significantly longer: \textbf{55.8 hours} vs. \textbf{16.0 hours} ($p < 0.001$; small effect size with Cliff's $\delta = 0.310$). The impact was most pronounced for OpenAI Codex (38.6 vs. 5.7 hours, $p < 0.001$) and refactoring tasks (51.2 vs. 15.5 hours, $p = 0.012$). 

Further, regression analysis\footnote{Detailed regression results are available in our \href{https://github.com/gjz78910/PR-MCI/blob/main/supplementary_material/regression_results.md}{\textcolor{blue}{supplementary material}}.} showed that PR-MCI remained significantly associated with lower acceptance and longer merge times ($p < 0.001$ for both) after controlling for code churn, files changed, task type, and agent, suggesting that PR-MCI has an independent association with outcomes beyond what can be explained by code complexity or task characteristics.

\begin{quotebox}
\noindent
\textbf{RQ3:} High-MCI PRs had \textbf{51.7\% lower acceptance} (28.3\% vs. 80.0\%) and took \textbf{3.5$\times$ longer to merge} (55.8 vs. 16.0 h), confirmed by statistical tests with confounders controlled, suggesting that PR-MCI is associated with reviewer trust and PR outcomes.
\end{quotebox}

\section{Discussion}
\label{sec:discussion}

\subsection{Actionable Insights for Developers}
Our findings demonstrate that even when AI-generated code is acceptable, PR descriptions may misstate scope or claim phantom changes, confirming why human oversight remains necessary for Agentic-PRs~\cite{watanabe2025use}. Therefore, we suggest: 

\begin{takeawaybox}
\textbf{For agent users:} (1) verify that agents actually commit changes, (2) avoid template-based prompts, and (3) refine prompts to guide agents toward more complete descriptions.
\end{takeawaybox}

For PR reviewers, we recommend three quick heuristic checks to identify PR-MCI without examining the diff in detail: 

\begin{takeawaybox}
\textbf{For PR reviewers:} (1) verify the diff is not empty, (2) {check if descriptions seem too brief for the amount of code changed}, and (3) {check for template markers} (e.g., [WIP], [Failed], or generic phrases like "I'm starting to work on it").
\end{takeawaybox}

\subsection{Actionable Insights for AI Tool Builders}
Our analysis reveals that high-MCI PRs take {3.5$\times$ longer to merge}, wasting around \textbf{40 hours per PR}. Even though only 1.7\% of PRs have high-MCI, they may affect trust in the AI-driven software development lifecycle \cite{gropler2025future}. Therefore, we recommend:
\begin{takeawaybox}
\textbf{For AI tool builders:} (1) implement automated PR-MCI verification mechanisms that can detect common types (e.g., Phantom Changes, Scope Understated, and Placeholder/Incomplete), and (2) integrate them into Agentic-PR generation pipelines to flag problematic descriptions before review, helping recover thousands of hours of wasted review effort.
\end{takeawaybox}

\subsection{Implications for SE 3.0 Research}
The emerging ``SE 3.0'' vision positions AI as active teammates in the software development lifecycle~\cite{aidev2025, gropler2025future}, where AI agents are expected not only to generate code, but also to communicate intent, justify changes, and support human decision-making in collaborative workflows. Our findings provide empirical evidence that current Agentic-PR descriptions may vary substantially in quality and can strongly affect review outcomes. 

Building on this insight, future research could leverage our 974 manually annotated PRs with 432 partial/misaligned cases to:

\begin{takeawaybox}
\textbf{For Researchers:} (1) investigate why agent communication abilities vary, (2) develop standardized evaluation metrics for PR description quality, and (3) explore how Agentic-PR generation can be improved through better prompt engineering~\cite{gong2025tuning, huang2024effilearner}, fine-tuning~\cite{huang2024swiftcoder, gong2025language}, or reinforcement learning~\cite{gong2025enhancing, turker2024accelerating, thomas2024muprl}.
\end{takeawaybox}

\subsection{Threats to Validity}
This study has several limitations that should be considered when interpreting the results. (1) \textbf{Construct validity:} Our PR-MCI metric is a heuristic proxy for semantic consistency and may miss subtle mismatches or penalize concise but accurate descriptions. (2) \textbf{Threshold calibration:} The decision threshold is calibrated on a validation set, which weakens absolute prevalence claims despite observed stability under cross-validation. (3) \textbf{External validity:} Dataset filtering choices prioritize scale over strict controls and may affect generalizability. (4) \textbf{Sample size imbalance:} Agent sample sizes are imbalanced (e.g., Claude Code has only 220 PRs), limiting the reliability of agent-specific conclusions for smaller samples. (5) \textbf{Taxonomy completeness:} The taxonomy is derived from a limited subset of PRs (432 partial/misaligned cases) and may not capture all inconsistency types. (6) \textbf{Causal inference:} Our analysis is observational and does not establish causal relationships between PR-MCI and review outcomes. (7) \textbf{Ethical considerations:}
This study analyzes only publicly available GitHub data from repositories under
permissive licenses (MIT or Apache~2.0) and reported findings in aggregate to
avoid identifying individuals or projects.

\section{Conclusion}
\label{sec:conclusion}

We used PR-MCI to quantify alignment between AI-generated PR descriptions and code changes, analyzing 23,247 PRs from five AI agents. Our findings reveal that unreliable PR descriptions (high PR-MCI) are associated with significantly lower PR acceptance rates and longer merge times, indicating that improving the consistency and accuracy of AI-generated PR descriptions is essential for trustworthy human-AI collaboration.

\vspace{0.2cm}
\noindent \textbf{Data Availability.}
All data, scripts, results, and supplementary materials are available at 
\href{https://github.com/gjz78910/PR-MCI}{\textcolor{blue}{https://github.com/gjz78910/PR-MCI}}.

\vspace{0.2cm}
\noindent \textbf{Acknowledgment.}
This work has been supported by the ITEA GENIUS grant (project number 23026).

\balance
\bibliographystyle{ACM-Reference-Format}
\bibliography{references}

\end{document}